\documentclass{SCIYA2017enOL}
\online
\begin{document}

\ensubject{fdsfd}

\ArticleType{ARTICLES}
\Year{2017}
\Month{January}%
\Vol{60}
\No{1}
\BeginPage{1} %
\DOI{10.1007/s11425-000-0000-0}
\ReceiveDate{December 6, 2019}
\AcceptDate{January 16, 2020}

\title[]{Noise control and utility: from regulatory network to spatial patterning}
{Noise control and utility: from regulatory network to spatial patterning}

\author[1,2,$\ast$]{Qing Nie}{qnie@uci.edu}
\author[3]{Lingxia Qiao}{}
\author[1]{Yuchi Qiu}{}
\author[3,4,$\ast$]{Lei Zhang}{zhangl@math.pku.edu.cn}
\author[4]{Wei Zhao}{}

\AuthorMark{Nie Q}

\AuthorCitation{Nie Q, Qiao L X, Qiu Y C}


\address[1]{Department of Mathematics, University of California Irvine, Irvine, CA \rm{92697}, USA}
\address[2]{Department of Developmental and Cell Biology, NSF-Simons Center for Multiscale Cell Fate Research, \\
University of California Irvine, Irvine, CA \rm{92697}, USA}
\address[3]{Beijing International Center for Mathematical Research, Peking University, Beijing \rm{100871}, China}
\address[4]{Center for Quantitative Biology, Peking University, Beijing \rm{100871}, China}


\abstract{Stochasticity (or noise) at cellular and molecular levels has been observed extensively as a universal feature for living systems. However, how living systems deal with noise while performing desirable biological functions remains a major mystery. Regulatory network configurations, such as their topology and timescale, are shown to be critical in attenuating noise, and noise is also found to facilitate cell fate decision. Here we review major recent findings on noise attenuation through regulatory control, the benefit of noise via noise-induced cellular plasticity during developmental patterning, and summarize key principles underlying noise control.}



\maketitle

\section{Introduction}
One fundamental task in biology is to understand how cells perform complex functions accurately and robustly in face of inevitable biological noise. Due to the fluctuating environments and inherently stochastic biochemical reactions, many cellular processes including signal transduction and gene expression operate in a substantially noisy way \cite{2002Elowitz,2002Swain}.  As a result, the temporal fluctuations or cell-cell variability of a molecule in its concentration, activity, modification form, or cellular localization, often referred as ``noise", are observed \cite{2010Eldar,2000Elowitz}

Over the last decade, noise in gene circuits has been observed, dissected and  analyzed to understand how cell executes complex  functions. Multiple biochemical processes can contribute to the noise in specific cellular behavior, and reversely, the observed noise can be decomposed into different components based on the sources of noise, providing a deep insight into the roles of each component on the cellular behavior \cite{2002Elowitz,2004Raser,2015Liu,2017Colin}. Besides, quantitative analysis of the observed noise supplies the information about the biological network, such as the network topology and the regulation interaction \cite{2019Shi, 2012Munsky}. Furthermore, increasing studies investigate how to control or utilize noise in living systems \cite{2011Sartori,2017Colin, 2010McCullagh, 2005Lerner, 2005Forger, 2004Mihalcescu, 2015Sartori, 2008Ghim,2011Sartori,2015Wang,2012Zhang}. While noise may induce heterogeneity within the cell population, contributing to diversity in cell fate choice \cite{2015Ge,2008Acar,2005Kussell}, it usually causes uncertainty in information transmission in the cell and impairs robustness of cellular functions, which is one of the  main reasons for the difficulty in  robust circuit functions \cite{2000Elowitz, 2016Potvin}. A natural question at hand is how cells deal with noise effectively.

Since cellular functions, such as bistability, oscillation and adaptation, have been found to link to regulatory network architectures \cite{2013Lim}, the network property is naturally important in noise control \cite{2000Becskei, 2002Houchmandzadeh,2005Hooshangi, 2005Brandman, 2008Hornung, 2010Wang, 2012Zhang, 2013Chen, 2018Yu}. What are basic constraints on the regulatory networks for noise attenuation? How noise is controlled in function-specific systems, such as adaptive systems or oscillatory systems?
In addition, can noise be utilized to achieve specific biological functions? How does noise affect spatial organization and morphogen-mediated  patterning? How can a precise and robust readout be generated from the noisy spatial morphogen gradient?

In this work, we first review major noise attenuation mechanisms in regulatory networks, and  then explore key strategies to combat noise in morphogens during spatial patterning. We conclude by summarizing the major mechanisms in noise control.

\section{Noise attenuation mechanisms in regulatory networks}
To study how cell regulatory network functions robustly in the presence of noise, many efforts have been made through experiments and theoretical approaches. By encoding a fluorescent protein as a tag of a protein of interest, the expression in each single cell at any observed time is quantified through fluorescence intensity. Therefore, the expression noise with time or in a population of clonal cells can be measured by using the coefficient of variation (CV, i.e., standard derivation over the mean) or Fano factor (i.e., variance over the mean). From a theoretical or computational point of view, the molecular species are modelled as discrete random variables, which are produced or hydrolyzed when reaction events (such as translation, transcription and protein-protein binding) occur. Reductions of the model, including assuming molecular species as continuous variables, merging multi-steps to one reaction and simplifying the noise term, are also adopted according to the scale of the question. By analyzing or simulating the model, one can obtain the analytical expression or numerical value of noise.

In the last two decades, noise attenuation mechanisms have been largely explored in diverse cellular networks, such as the chemotaxis pathway in \emph{Escherichia coli} \cite{2011Sartori,2017Colin},  MAP (mitogen-activated protein) kinase pathway \cite{2010McCullagh,2005Lerner}, and the pathway controlling mammalian circadian clock \cite{2005Forger, 2004Mihalcescu}. General principles that are applicable for most biological networks have been identified \cite{2015Sartori, 2008Ghim,2011Sartori,2015Wang}, and among them the network topology and timescale emerge as  two important factors for noise attenuation. Since the cell performs multi-functions simultaneously, subsequent research is also aimed at understanding how to handle the cellular noise in function-specific systems.

\begin{figure}[h!]
\centering
\includegraphics[width=15cm]{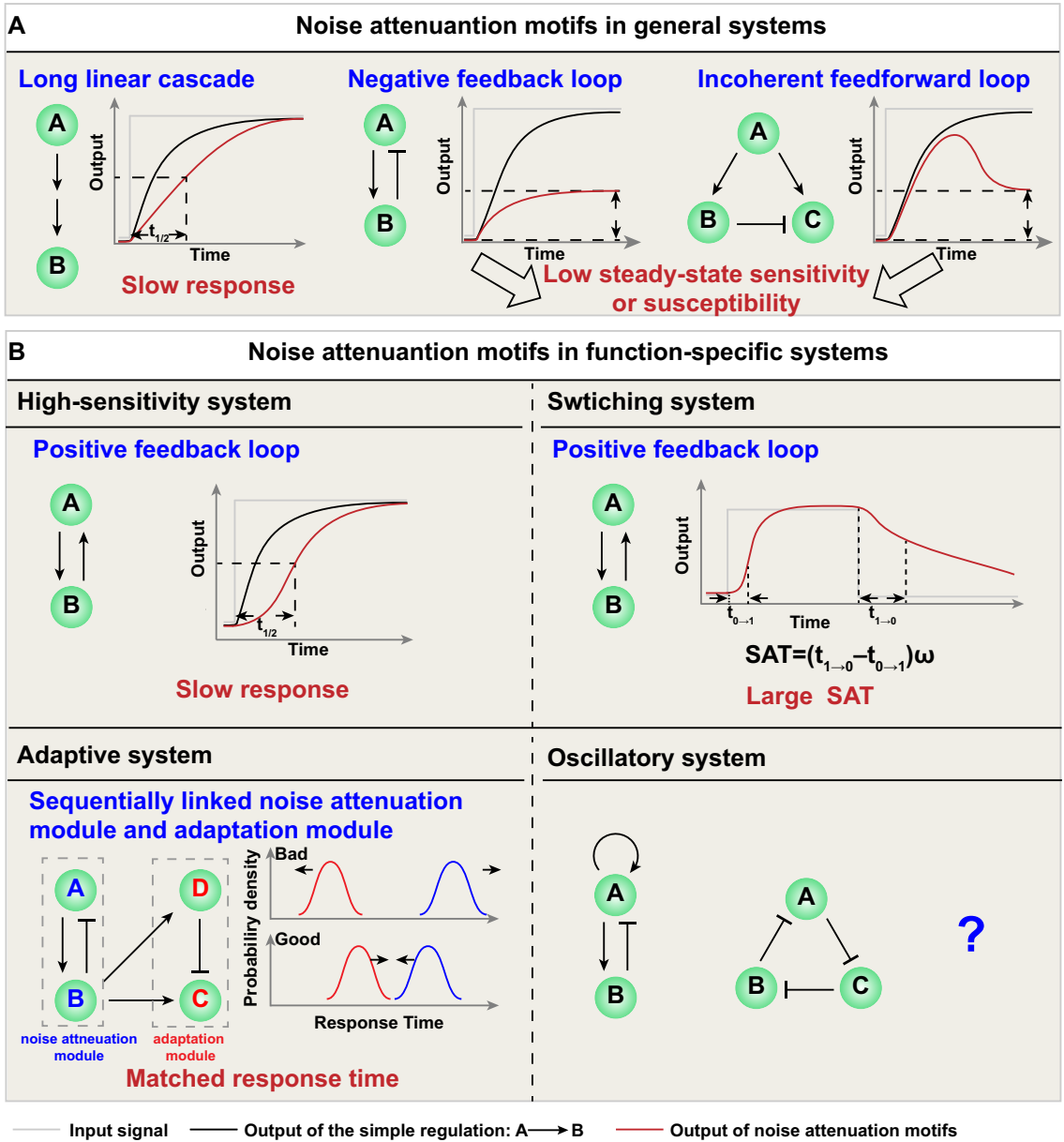}
\caption{\textbf{Noise attenuation motifs and the corresponding mechanisms. (A-B) Three generic noise attenuation motifs that are applicable for most biological systems (A) and noise attenuation motifs in four function-specific systems (B). The noise attenuation motifs in general systems include long linear cascade, negative feedback loop and incoherent feedforward loop. The four function-specific systems are high-sensitivity system, switching system, adaptive system and oscillatory system. For all network topologies, node A receives the input signal, while node B is the output node except that the output node for incoherent feedforward loop or adaptation module is node C. Dashed lines indicate quantities related to noise buffering capability, including the response time $t_{1/2}$ (defined by the time of output to reach halfway of the steady state), the output change magnitude and SAT (refer to Eq. (\ref{SAT}) for the definition). In the adaptive system, distributions of response time for the noise attenuation module and the adaptation module are schematically plotted in blue and red, respectively. } }
\label{fig_motif}
\end{figure}

\subsection{The role of network topology and timescale}

It is widely accepted that structures determine functions. Likewise, the topology of the regulatory network may be crucial for noise buffering capability. Noise attenuation motifs and the corresponding mechanisms have been summarized in Figure \ref{fig_motif}. While Figure \ref{fig_motif}A presents three generic noise attenuation motifs that are applicable for most biological systems, Figure \ref{fig_motif}B illustrates the noise attenuation motifs in four function-specific systems. 

Linear cascade acts as a low-pass filter by attenuating high-frequency fluctuation in the input signal \cite{2005Hooshangi}. This property is due to time delay of reactions (Figure \ref{fig_motif}A) and many biological networks also possess it. However, the gene expression noise is not only determined by the noise in the input signal level but also the intrinsic fluctuations and variations in the reaction rates \cite{2005Pedraza}. Compared with the short cascade, the long cascade exhibits better capability of filtering fluctuations in the input signal level while at the same time accumulating more inherent noise. In order to achieve small output noise, the length of cascades needs to be optimized by considering the trade-off between the noise buffered by time lags and noise accumulated in the pathway \cite{2002Thattai}.

Negative feedback (NF) is a common noise-attenuating regulatory network. The principle of NF is that an increase (decrease) of one species can lead to repression (activation) of itself through one or multi-step chemical reactions. Its dynamic behavior helps to suppress fluctuations, thus contributing to the stability of the species \cite{2000Smolen}. This effect can be also demonstrated by the comparison of negative autoregulation, positive autoregulation and simple regulation in the synthetic gene circuit. Adding the negative feedback in simply regulated genes reduces the cell-cell variation of protein levels \cite{2000Becskei}. In contrast, positive autoregulation increases the cell-cell variation compared with negative autoregulation \cite{2007Alon}. Strong positive autoregulation can even induce the bimodal distribution of protein levels.

In fact, the noise buffering capability of NF arises from the system's low steady-state sensitivity (or susceptibility) \cite{2008Hornung} and incoherent feedforward loops adopt similar mechanism (Figure \ref{fig_motif}A). However, when the sensitivity to input signal is required, positive feedback (PF) shows superiority in noise attenuation (Figure \ref{fig_motif}B). In most cases, sensitivity is defined as the steady-state or transient response to a small change of the input signal, leading to two versions of sensitivity:  steady-state sensitivity (or susceptibility)
\begin{equation}Sensitivity_{steady}=|\frac{(O_2-O_1)/O_1}{(I_2-I_1)/I_1}|\end{equation}
and transient sensitivity
\begin{equation}Sensitivity_{transient}=|\frac{(O_{peak}-O_1)/O_1}{(I_2-I_1)/I_1}|\end{equation}
where $O_1$, $O_{peak}$ and $O_2$ are the initial output steady state, transient peak value, final output steady state respectively when the input signal changes from $I_1$ to $I_2$. The steady-state sensitivity can be rewritten as $|\mathrm{d}(\mathrm{ln}  \ O_1)/\mathrm{d}(\mathrm{ln}  \  I_1)|$ when $I_2-I_1$ is infinitesimally small. By numerically analyzing networks consisting of an input node and a two-node network, the output noise of PF is found smaller than that of NF for a given steady-state susceptibility to long-term changes in input signals \cite{2008Hornung}. This advantage also holds when transient sensitivity is used \cite{2019Qiao}. Furthermore, if we consider the condition when the change of the input signal is large, i.e., the input signal is removed or added, PF still exhibits lower noise in both ON and OFF states compared with NF in switching systems \cite{2013Chen}. The main reason resides in the fact that PF usually exhibits longer response time than NF. At the early stage of input change, the output of PF usually increases slowly while that of NF rises rapidly. As a consequence, PF needs more time to reach the halfway to the steady state, thus slowing down the dynamics. Because the prolonged response time provides good time-averaging of fluctuations, PF has better noise buffering capability than NF.

The fact that longer linear cascade provides better time-averaging for noise buffering and the fact that the different noise buffering capabilities of PF and NF are related to their different timing of response, both indicate the important role of timescale in noise attenuation. Here, timescale represents how fast the system responds to the input change.

An intrinsic quantity termed the signed activation time (SAT), directly measured through the system's dynamic properties in the absence of the noise, has been identified to affect the noise buffering capability \cite{2010Wang}. The SAT is defined as the difference between the deactivation and activation timescales relative to the input noise timescale:
\begin{equation}SAT=(t_{1\rightarrow0}-t_{0\rightarrow1})\cdot \omega \label{SAT},\end{equation}
where $\omega$ is the frequency of input noise, and $t_{1\rightarrow0}$ and $t_{0\rightarrow1}$ denote the deactivation timescale and the activation timescale respectively. The output noise is quantified by the noise amplification rate, defined as the ratio of output CV to the input CV:
\begin{equation}NAR=\frac{{\rm std}(O)/{\rm mean}(O)}{{\rm std}(I)/{\rm mean}(I)}.\end{equation}
An inverse relationship between SAT and the noise amplification rate has been found when the switching system settles in the ON state \cite{2010Wang}. It means that switching systems with both fast activation and slow deactivation, i.e., large SAT, are advantageous in noise attenuation. An intuitive explanation is illustrated as follows: fast activation helps the output to rebound rapidly from the OFF state while slow deactivation prevents the output from falling into the OFF state, thus arresting the output at the ON state and reducing fluctuations.

The inverse relation between SAT and noise amplification rate is validated by theoretical and numerical studies in systems of two positive feedback loops, systems of one positive feedback loop and one negative feedback loop, and six complex regulatory systems \cite{2010Wang}.The single positive feedback loop (Figure \ref{fig_motif}B), incorporating a mutual activation loop of $A$ and $B$, can be used to theoretically illustrate the inverse relation. The system can be modeled by the following ordinary differential equations:
\begin{equation}
\begin{aligned}
\frac{db}{dt}&=k_1a(1-b)-k_2b+k_3\\
\frac{da}{dt}&=(k_bsb(1-a)-a+k_4)\tau_a,\\
\end{aligned}
\end{equation}
where $a$ and the output $b$  are the normalized concentrations of $A$ and $B$, respectively. The normalized input signal $s$ ranges between the OFF state (i.e., $s=0$) and the ON state (i.e., $s=1$), and varies over time with the frequency $\omega$. A general form of $s$ can be taken as follows: $s=0$ when $t \leq T_0$ and $s=1+\xi_{\omega}(t)$ when $t >T_0$, where $T_0$ is the time point when the system changes from the OFF state to the ON state and $\xi_{\omega}(t)$ is the noise term. Typically, $\xi_{\omega}(t)$ is constructed by dividing the time interval into subintervals with length $1/\omega$ and then applying independent random numbers from a uniform distribution in $[-1,1]$ to $\xi_{\omega}(t)$ on each subinterval. 
$k_b$, $k_1$,  $k_2$, $k_3$ and $k_4$ are kinetic parameters and the inverse of $\tau_a$ represents the timescale of $A$. On the one hand, the fluctuation dissipation theorem (FDT) is used to calculate the noise amplification rate \cite{2008Hornung,2004Paulsson,2010Wang}. When $\tau_a\ll 1$ and $\tau_a/ \omega\ll 1$, the noise amplification rate is
\begin{equation}
NAR\approx\sqrt{\frac{\tau_a/\omega}{\left \langle s\right \rangle(K_bk_b-1)(K_b+1)\frac{k_b}{k_b+1}}},
\end{equation}
where $K_b=k_1/k_2$ is the association constant. Because $K_bk_b-1>0$ is always true for the switching system, NAR is an increasing function of $\tau_a$ and decreasing function of $K_b$ and $k_b$. On the other hand, the deactivation and activation timescales are derived by linearizing the noise-free system around the ON state and the OFF state, respectively. In order to achieve slow deactivation time, $\tau_a$ needs to be small enough. With a fixed small $\tau_a$, the deactivation timescale is characterized by
\begin{equation}
(K_b+1)\frac{k_b}{k_b+1}.
\end{equation}
So, increasing $K_b$ or $k_b$ contributes to slow deactivation time. Similarly, under the assumption of small fixed $\tau_a$, the activation timescale can be accelerated by decreasing the following quantity
 \begin{equation}
\frac{1}{K_b+1}\left(1+\frac{1}{k_b}\right)
\end{equation}
 and increasing $K_b$ or $k_b$ also speeds up the activation process. As a result, the noise amplification rate is supposed to depend negatively on $t_{1\rightarrow0}-t_{0\rightarrow1}$. Based on the equation of the noise amplification rate, the frequency of input signal $\omega$ negatively affects the noise amplification rate. These facts, taken together, theoretically indicate the inverse relation of noise amplification rate and $(t_{1\rightarrow0}-t_{0\rightarrow1})\cdot \omega$ in the single positive feedback loop. Furthermore, SAT in the single positive feedback loop links the noise amplification rate and the sensitivity to the input signal:
  \begin{equation}
NAR\approx\frac{s_{on}}{\sqrt{SAT}},
\end{equation}
where  $s_{on}$ denotes the steady-state sensitivity in the ON state \cite{2013Chen}.

The SAT can only explain the system's noise buffering capability at the ON state, but not at the OFF state. To address this issue, researchers introduced a modified version of SAT --input-associated SAT (iSAT) to capture the noise buffering capability at the OFF state in the switching system \cite{2013Chen}. The definition of iSAT at the OFF state is the same as the SAT except that the input change is relatively small. However, at the OFF state, the iSAT has a positive relationship with the noise amplification rate, which is validated in several feedback systems. It means that slow activation and fast deactivation help to fix the output at the OFF state. Thus, there is a trade-off between the noise buffering capability at the ON state and at the OFF state. Nevertheless, an exploration of 33 three-node circuits identifies five network topologies capable of producing both high SAT at the ON state and low iSAT at the OFF state. Interestingly, all these five network topologies contain the mutual activation motif.

Nevertheless, the SAT exhibits topology dependence.  A large SAT can be obtained in the single positive feedback loop (Figure \ref{fig_motif}B), while an additional positive feedback loop tends to further increase the SAT by accelerating the activation and keeping the deactivation slow. This is consistent with the previous work showing the superior noise attenuation at the ON state of the interlinked fast and slow positive feedback loops \cite{2005Brandman}.

The interplay between network topology and noise buffering capability can be partially explained by the timescale. 
For example, linear cascades have the effect of time delay, and cascade with long length is good at filtering input signal noise due to enough time delay; positive feedback responses more slowly than negative feedback when the same sensitivity is required, and thus leads to better noise buffering capability; interlinked fast and slow positive feedback loops, where the output turns on rapidly and turns off slowly, have been identified to maintain a robust high state \cite{2005Brandman}. In summary, timescale may be a bridge linking network topology to noise buffering capability.

\subsection{Rules to control noise in adaptive systems}

While early studies identified several general principles for noise attenuation, recent explorations of noise attenuation mechanism have been extended to diverse signaling systems such as those executing adaptation, oscillation and cell fate decision \cite{2013Ji, 2018Hansen, 2011Fritsche, 2016Potvin, 2017An,2015Wang}. The challenge for the latter is to coordinate multiple functions together, considering different functions could interfere each other. For example, during the lineage commitment process, the noise helps to induce the cell fate selection, and then is suppressed to stabilize the cell fate \cite{2012Zhang, 2018Hansen}.

Adaptation, the system's ability to sense the change of stimulus and finally go back to the pre-stimulated level, widely exits in biological systems such as bacteria and amoeba \cite{1947Bonner,1999Alon}. This function enlarges the sensing range, resists fluctuating environments and shuts down the signaling response timely. A systematic search of all three-node networks demonstrates that a negative feedback loop with a buffering node (NFBLB) and an incoherent feedforward loop with a proportioner node (IFFLP) are two core motifs in order to achieve the adaption \cite{2009Ma}. Since adaptive systems are always executed with noise, the accuracy of the adaptation is reduced and thus how to achieve accurate adaptation is of interest. Most signaling pathways need external metabolic energy, so they are operated out of equilibrium and do not follow the linear relationship between noise and sensitivity (fluctuation dissipation theorem). As a result, it is possible for signaling networks to achieve the aim: high sensitivity (i.e. transient sensitivity)  to stimulus and low noise simultaneously \cite{2015Sartori}. For different adaptive systems, negative feedback loop is more sensitive to the change of the signal than incoherent feedforward loop for a given output noise level caused by the intrinsic stochasiticity of chemical reactions \cite{2015Shankar}. In terms of timescale, adaptive systems act as a bandpass filter with respect to the input signal: high-frequency extrinsic noise is filtered by the effect of time-averaging and low-frequency extrinsic noise is buffered by the adaptation dynamics \cite{2011Sartori}.

Whereas early work studied the mechanism of achieving noise attenuation and adaptation simultaneously on a case-by-case basis, there exists the general design principle from the bottom up \cite{2019Qiao}. Since adaptation requires at least three nodes (i.e., an input node, an output node and an intermediate node), the network topologies achieving dual function of adaptation and noise attenuation are supposed to be those with three or more nodes. Unfortunately, an exhausted search of all three-node networks shows that the trade-off between noise attenuation and adaptation hinders the achievement of this dual function in three-node networks. Given the important role of timescale, strategies of tuning the timescale are investigated. Taking the enzymatic regulatory IFFLP (the incoherent feedforward loop in Figure \ref{fig_motif}A) as an example, we modeled the system and studied the effect of timescale. In this system, input (I) activates enzyme A and then the active form of enzyme A catalyzes the conversion from inactive forms of enzyme B and C to their active forms, while the active form of enzyme B catalyzes the conversion from the active form of enzyme C to its inactive form. The corresponding dynamics can be described by a set of differential equations:
\begin{equation}
\begin{aligned}
\tau_A \frac{dA}{dt}&=f_{A}(I,A) \triangleq Ik_{IA}\frac{1-A}{1-A+K_{IA}}-F_Ak_{F_AA}\frac{A}{A+K_{F_AA}}  \\
\tau_B\frac{dB}{dt}&=f_{B}(A,B) \triangleq Ak_{AB}\frac{1-B}{1-B+K_{AB}}-F_Bk_{F_BB}\frac{B}{B+K_{F_BB}}  \\
\tau_C\frac{dC}{dt}&=f_{C}(A,B,C) \triangleq Ak_{AC}\frac{1-C}{1-C+K_{AC}}-Bk_{BC}\frac{C}{C+K_{BC}}  
\end{aligned}
\end{equation}
where $I$ is the input signal as a function of time $t$. Dynamical variables $A$, $B$ and $C$ are the concentrations of the active enzyme A, B and C, respectively. By normalizing the total concentration of each enzyme as 1, $1-A$, $1-B$ and $1-C$ represent concentrations of the three enzymes in inactive form. $f_i,i=A,B,C$ denotes the reaction rate of the active enzyme. $F_A$ and $F_B$ (set to be constant) are concentrations of the basal deactivating enzymes of $A$ and $B$ respectively. $K$'s and $k$'s are Michaelis-Menten constants and catalytic rate constants, respectively. $\tau_A$, $\tau_B$ and $\tau_C$ represent the timescales of input node A, intermediate node B and output node C, respectively. When measuring the sensitivity and precision, the input signal $I$ is a step function of $t$. When calculating the noise buffering capability, a zero-mean noise term whose autocorrelation function dies out exponentially with timescale $\tau_0$ (the inverse of the frequency $\omega$) is added to input $I$. By assuming the same dynamics of the input and the node A and linearizing the system,  the sensitivity and NAR for enzymatic regulatory IFFLP satisfying perfect adaptation can be approximated as:
\begin{equation}
Sensitivity_{transient}=\frac{k_4}{k_6}(\frac{k_1\tau_C}{k_6\tau_B})^{\frac{\frac{k_1\tau_C}{k_6\tau_B}}{1-\frac{k_1\tau_C}{k_6\tau_B}}}
\end{equation}
and
\begin{equation}
NAR=\sqrt{\frac{{k_4^2}({\frac{\tau_B}{\tau_C}})^2\omega}{(k_6\frac{\tau_B}{\tau_C}+k_1)(k_6\frac{\tau_B}{\tau_C}+\omega\tau_B)(\frac{k_1}{\tau_B}+\omega)}}
\end{equation}
respectively. Here, $k_1=\frac{k_{F_BB}F_B}{K_{F_BB}}$, $k_4=\frac{k_{BC}\langle B \rangle}{\langle C \rangle+K_{BC}}$ and $k_6=\big[\frac{K_{AC}\langle C \rangle}{(1-\langle C \rangle+K_{AC})(1-\langle C \rangle)}+\frac{K_{BC}}{\langle C \rangle+K_{BC}}\big]k_4$, where $\langle \cdots \rangle$ denotes the steady-state value. With fixed $\frac{\tau_C}{\tau_B}$, it can be seen that sensitivity remains constant and NAR is a decreasing function of $\tau_B$ (or $\tau_C) $ if $\tau_B\tau_C\geqslant \frac{{k_1}{k_6}}{\omega^2}$ \cite{2019Qiao}. Together with numerical simulations, the analytic derivation of NAR and sensitivity confirms that increasing both the timescale of the output node and of the intermediate node can dramatically decrease NAR while maintaining sensitivity. Nevertheless, this strategy introduces prolonged adaption time and unrealistic parameter ranges. It seems that three-node networks are difficult to achieve the dual function of perfect adaptation and noise attenuation.

By contrast, four-node networks are able to achieve the dual function by sequentially combining adaptation module and noise attenuation module (Figure \ref{fig_motif}B). However, placing the noise attenuation module upstream the adaptation module (N-A) exhibits better performance of the dual function than combining modules in the reverse order (A-N). Besides, the performance of achieving the dual function in the sequentially combined four-node network is affected by the topology of the functional module. The main reason is that different topologies of the functional module possess different response time and thus cause distinct abilities to maintain high sensitivity after combination, which greatly affects the performance of the dual function in sequentially combined four-node networks as a consequence. Typically, the noise attenuation module has a longer response time than the adaptation module under the same parameter space, because noise attenuation needs time averaging. However, a matched response time of the two modules is advantageous in combining the high-performance dual function network. For N-A networks, a relatively fast response of the noise attenuation module contributes to high sensitivity of the downstream output. For the A-N networks, the upstream adaption dynamics with relatively slow response can avoid being largely filtered by the noise attenuation module. In summary, module combination and timescale matching together overcome the trade-off between noise attenuation and adaptation. Finally, the examination of seven biological systems suggests that in many adaptive systems a positive feedback loop or negative feedback loop is coupled with the adaptation module, creating an expanded dual-function module.

\subsection{Mystery of noise in oscillatory systems}
Oscillations are ubiquitous in a broad range of biological processes, including circadian clocks \cite{1998Ouyang}, NF-$\kappa$B dynamics in immune response \cite{2010Tay}, and vertebrate somitogenesis \cite{2012Oates}. These oscillations allow organisms to know the time and govern many biological functions: the daily rhythm enhances the growth of many organisms; NF-$\kappa$B is an important transcription factor in inflammatory response; the segmentation clock is thought to regulate the formation of somites. To design a biochemical oscillator, negative feedback, time delay and nonlinearity are required. The repressilator (the second network topology in the oscillatory system in Figure \ref{fig_motif}B), a circle composed of three species each of which acts as a repressor to negatively regulate the next one, is a relatively simple system to achieve oscillation. The model of this system with transcriptional regulation is as follows \cite{2000Elowitz}:
\begin{equation}
\begin{array}{l}
\frac{dm_i}{dt}=\alpha_0+\frac{\alpha_i}{1+(\frac{p_j}{K_{ji}})^n}-d_im_i,(i,j)=\{(A,B),(B,C),(C,A)\}, \\
\frac{dp_i}{dt}=\beta_i m_i-\gamma_ip_i,i=A,B,C,
\end{array}
\label{eq_rep}
\end{equation}
where $m_i$ and $p_i$ are concentrations of the mRNA and the protein, respectively. $d_i$ and $\gamma_i$ are decay rates; $\beta_i$ is translation rate; $\alpha_0$ is the basal transcription rate while $(\alpha_0+\alpha_i)$ is the transcription rate in the absence of its repressor; $K_{ji}$ is the repression coefficient. Under proper parameter configurations, the system described by Eq. (\ref{eq_rep}) can oscillate regularly. However, biological systems are subject to substantial noise due to the fluctuating environment and stochasticity of chemical reactions. The presence of the noise leads to the variability of the oscillation features such as period and amplitude, and thus impairs the accuracy of oscillation. By introducing noise term into the deterministic differential equations (i.e., stochastic differential equations) or directly simulating the system with Gillespie algorithm \cite{1977Gillespie}, one can obtain stochastic trajectories for observed variable $x$. The accuracy of oscillation can be captured by the autocorrelation function $C(t)$ for the observed variable $x$, defined as
\begin{equation}
 C(t)=\frac{\langle(x(t+s)-\langle x \rangle)(x(s)-\langle x \rangle)\rangle_s}{\langle x^2 \rangle-{\langle x \rangle}^2}
\end{equation}
where $\langle \cdots\rangle$ represents the average and $s$ is a time variable \cite{2015Cao}. The autocorrelation function $C(t)$ usually follows a damped oscillation and the fast decay of $C(t)$ indicates that the oscillation system is not robust owning to little memory. The peak-time diffusion constant $D$ is another quantity \cite{2015Cao}, which is defined as the ratio of peak time variance $\sigma^2$ to the average peak time $t$:
\begin{equation}D=\sigma^2/t.\end{equation} 
Besides, the coefficient of variation for period or amplitude is also used to measure the oscillation accuracy \cite{2015Veliz}.

Understanding how biological systems achieve accurate oscillations remains a challenging work. Efforts have been made to control the oscillation dynamics in the noisy environment \cite{2000Barkai ,2002Vilar, 2004Stelling, 2005Forger, 2008Tsai, 2008Stricker, 2015Cao, 2016Potvin, 2018Fei}. The circadian clock driven by only two key elements is slightly affected by the intrinsic fluctuations \cite{2002Vilar}. In an oscillator incorporating interlinked positive feedback and negative feedback loop, strong positive feedback strength confers robust amplitude with respect to fluctuations of the synthesis rate \cite{2008Tsai}. Removing existing parts in the synthetic repressilator highly improves the regularity of oscillations \cite{2016Potvin}. Besides the important role of kinetic parameters in the accuracy of oscillations, these studies also indicate the effect of network topology on noise suppression in the oscillatory system \cite{2008Stricker, 2015Chen}. While one negative feedback loop can generate the oscillation, adding a positive feedback loop seems to provide better accuracy of  oscillation \cite{2015Chen}. Nevertheless, as so far, it seems that a  general network design principle for noise control in oscillatory systems is still lacking (Figure \ref{fig_motif}B). For example, whether above strategies to improve the accuracy of oscillation can be extended to other oscillatory systems is still unclear. Also, the underlying mechanism is to be further investigated.

\section{Stochastic dynamics in spatial patterning}
In developmental biology, one goal is to understand how cells execute specific cellular processes precisely based on their locations to form a robust spatial organization. A diffusive chemical, morphogen, having non-uniform distribution in space, plays an essential role in governing the pattern formation \cite{a1, a2}. The Turing model \cite{a3} and the morphogen gradient model \cite{a4} are two major models for explaining morphogen-mediated pattern formations. The Turing model gives rise to spatial pattern by a self-organized mechanism through interactions between two morphogens. In the morphogen gradient model, one morphogen forms a long-range gradient with a localized source and provides positional information to cells in a concentration-dependent manner. For examples, Turing models reveal intestinal crypts formations \cite{zhang2012reaction}, diverse feather shape \cite{a5}, digits pattern development \cite{a6} and periodic stripe formation in mammalian palate \cite{a7}. The morphogen gradient models are discovered in segmental pattern formations in vertebrate neural tube, drosophila embryo \cite{a8}, zebrafish hindbrain \cite{a9}, cell polarity \cite{2017Wang}, auxin transport \cite{2018Biha}, and skin stratification \cite{a11, a12}.

The simple morphogen gradient model sets a foundation for underlying patterning mechanisms in various systems. However, multiple sources, including environmental factors (e.g. temperature and nutrition), individual genetic differences and stochastic nature of biochemical processes, cause uncertainty in both morphogen and its downstream patterns. Therefore, several mechanisms will be discussed in this Section for reducing the uncertainty in different processes: morphogen formations and readouts, cell fate decisions and cell-cell interactions.

\begin{figure}[h!]
\centering
\includegraphics[width=15cm]{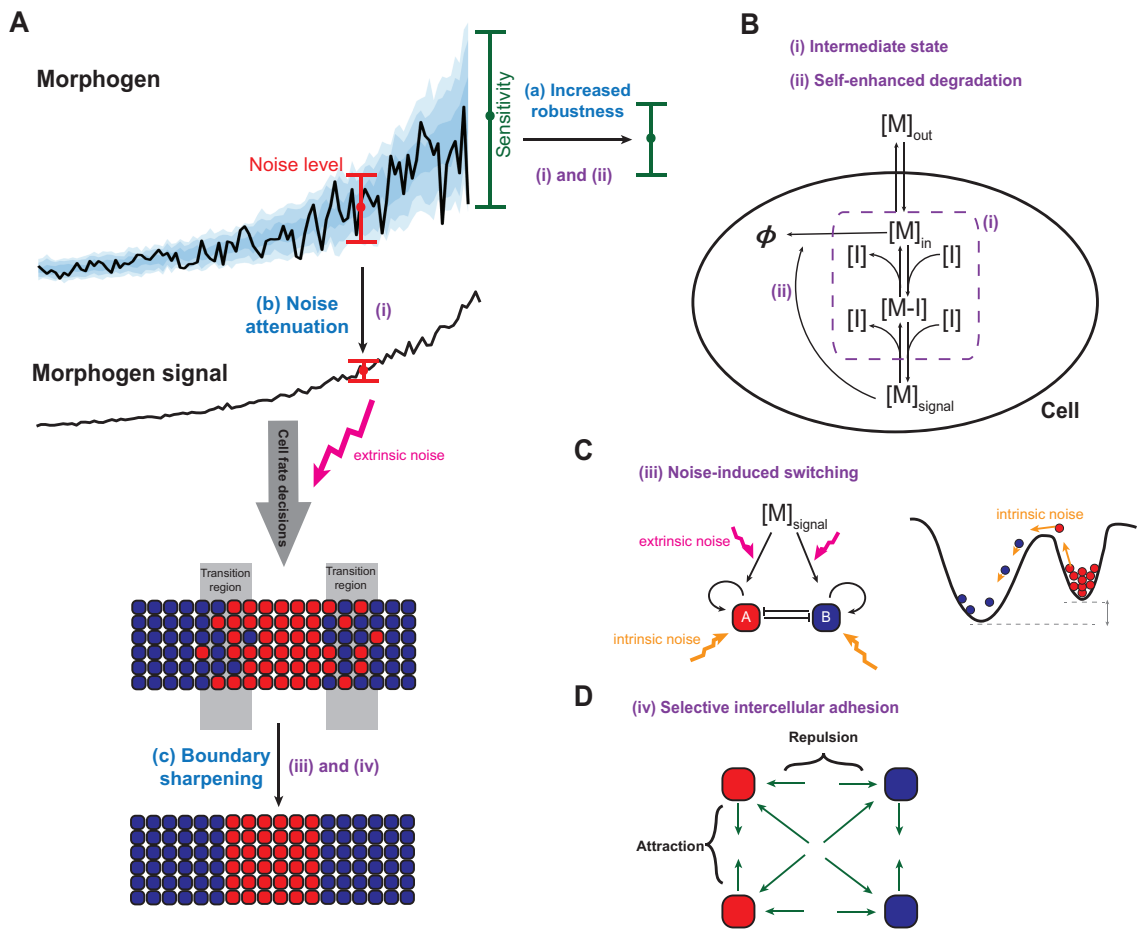}
\caption{\textbf{Noise attenuation in multiple processes: an example of zebrafish hindbrain segmental pattern formation.} (A) Illustration for segmental pattern formation from morphogen formation and readout to boundary sharpening. (B) Illustrations for (i) transcriptional network with intermediate state (i) in morphogen readout and self-enhanced degradation (ii) in morphogen formation; (C) Network of a genetic switch which allows noise-induced switching (left) and an illustration for explaining the noise-induced switching in the view of energy landscape (right); (D) Illustration for selective intercellular adhesion. $[M]_{\text{out}}$, $[M]_{\text{in}}$ and $[M]_{\text{signal}}$ are extracellular morphogen, intracellular morphogen and morphogen signals, respectively. $[I]$ is the intermediate state (binding protein in this example) in morphogen signaling cascade. $[M-I]$ is the complex formed by the binding between morphogen and intermediate state. }
\label{fig_morphogen}
\end{figure}
\subsection{Modeling morpohgens and their downstream signals}
First, we focus on morphogens formations and readout. Reaction--diffusion equations  are usually carried out to model morphogen formation in both extracellular and intracellular space \cite{a19}:
\begin{equation}
\label{eq_reac}
\begin{aligned}
&\frac{\partial[M]_{\text{out}}}{\partial t}=\overbrace{V(x)}^{\text{synthesis}}+D\overbrace{\Delta [M]_{\text{out}}}^{\text{diffusion}}\overbrace{-k[M]_{\text{out}}+k[M]_{\text{in}}}^{f}-\overbrace{d[M]_{\text{out}}}^{\text{degradation}}+\overbrace{\xi_{\text{out}}}^{\text{noise}},\\
&\frac{\partial[M]_{\text{in}}}{\partial t}=\overbrace{k[M]_{\text{out}}-k[M]_{\text{in}}}^{f}-\overbrace{S\left([M]_{\text{in}}\right)}^{\text{degradation}}+\overbrace{\xi_{\text{in}}}^{\text{noise}},
\end{aligned}
\end{equation}
where $[M]_{\text{out}}$ and $[M]_{\text{in}}$ are extracellular and intracellular morphogen concentration, respectively. Only extracellular morphogen has the ability to freely diffuse and the diffusion term is carried out. The spatial domain is defined on $[0,x_{\text{max}}]$. The production region of morphogen is a localized source and the synthesis term $V(x)$ is usually give by Heaviside unit step function \cite{lei2016mathematical}. For example, the source is defined on $\left[x_{\text{min}},x_{\text{max}}\right]$ and the synthesis term is given by:
\begin{equation}
\label{eq_syn}
V(x)=\frac{v}{x_{\text{max}}-x_{\text{min}}}H(x-x_{\text{min}}), \quad H(z)= \left\{
\begin{aligned}
0,\quad (z<0)\\
1,\quad (z\geq0)\\
\end{aligned}
\right.
\end{equation}
The term $f$ models the exchange of morphogen between intra and extra cellular forms. The degradation of extracellular morphogen is a linear term through the natural decay, while complex regulations lead to a nonlinear intracellular degradation of intracellular morphogen $S\left([M]_{\text{in}}\right)$. The noise term is usually given by white noise $\eta\frac{d\omega(t)}{dt}$ or colored noise.

After morphogen enters intracellular environment, morphogen transports to cell nucleus and binds with corresponding receptor to regulate transcriptions with help from complex signaling cascade. The final readout of morphogen $[M]_{\text{signal}}$ is the signal cells receive (Figure \ref{fig_morphogen}B). In Eq.\ (\ref{eq_reac}), the extracellular morphogen level is linearly proportional to the synthesis rate $v$ in Eq.\ (\ref{eq_syn}). The overall morphogen level is sensitive to the perturbations on synthesis rate. The noise terms in Eq.\ (\ref{eq_reac}) lead to noisy gradient distribution that disrupts precision of morphogen signals.

Many mechanisms, as we discussed in Section 2, such as complex transcriptional regulatory networks \cite{a8, a13} and negative feedback \cite{a14, a15} can also reduce sensitivity and attenuate noise in morphogen signals. Moreover, an intermediate state in the signaling cascade takes part in both reducing sensitivity and noise attenuation as well. One example of the intermediate state system is the retinoic acid (RA) signaling network in zebrafish hindbrain development. In this system, RA is synthesized at a local source and diffuses freely in the extracellular environment. The extracellular RA enters cell through cell membrane and the intracellular RA binds to a RA-binding protein (the intermediate state). The RA-binding protein transports RA to the nuclear receptor to form a complex that signals downstream gene expressions. All binding and unbinding with RA-binding protein are reversible (Figure \ref{fig_morphogen}B). With rapider binding and unbinding, better robustness to fluctuations on synthesis can be achieved in RA signals \cite{a16}, also lower noise level can be observed in RA signal without affecting the mean RA level \cite{a17, a18}. Different from non-spatial gene regulatory networks, a unique property in morphogen model is that diffusion spreads out morphogen over the space. The degradation controls the distance that morphogen can travel, and indeed, regulations on degradation improve the robustness as well. The self-enhanced degradation, where morphogen up-regulates its own degradation makes gradient level insensitive to synthesis and contributes robustness in both zebrafish hindbrain \cite{a19} and drosophila embryo pattern formations \cite{a20}.

\subsection{Utilizing gene expression noise to battle noise in morphogen}
The spatial coordination from morphogen specifies cell fate decisions, leading to tissue stratification and boundaries between regions with distinct identities. Despite of various noise attenutation mechanisms in morphogen formation and readout, signals that cells receive are still noisy, leading to uncertainty in fate decisions and fuzzy boundaries. Indeed, studies were carried out to investigate how cells make precise decisions despite receiving noisy signals. The temporal order of the gene appearance may vary in different system, despite they result in similar spatial alignment \cite{a21}. In the study of mandibular arch in zebrafish development, the temporal order of genes expression is found to modulate the response of the patterning network to noise and serves as a strategy for noise attenuation \cite{a22}.

Also, utilizing the intrinsic noise (e.g. noise in downstream gene expressions) to buffer the extrinsic noise (e.g. noise from signals) is a novel mechanism for noise attenuation. In gene regulatory networks, the intrinsic noise drives cell identity switching \cite{a23} and it is a survival strategy for cells in fluctuating environment \cite{2008Acar, a25}. Further studies on spatial pattern formations in multi-cellular organisms also reveal a proper range of intrinsic noise level is beneficial to obtain a finer spatial pattern. For example, in the development of zebrafish hindbrain, a stochastic gene expression model mimics a segmental pattern formation \cite{2012Zhang}. In this system, a genetic switch containing two genes and their interactions deploys two possible cell states. The auto-regulation generates a binary switch (ON or OFF) for each gene and the mutual inhibitions ensure at most one of two genes can be at ON state in each cell. The extracellular signal, morphogen, acting as an input on the genetic switch, determines the number of cell states. With high or low level signal, there is only a single state. Within a medium range of signaling level, called transition region, bistability appears. In the transition region, the cell fate highly depends on initial gene expressions in deterministic system and extrinsic noise (e.g. noise on morphogen signaling) gives rise to variability on initial gene expression leading to the co-existence of two cell states in a salt-and-pepper arrangement. These two states have different potentials from the view of energy landscape. The intrinsic noise, gene expression noise in the genetic switch, drives cells in the state with high potential switching to the state with low potential. This noise-induced boundary sharpening process leads to a homogeneous cells distribution with low potential state in transition region (Figure \ref{fig_morphogen}A and \ref{fig_morphogen}C). Many theoretical works have devoted to study the boundary sharpening process. By utilizing energy landscape, functions of each gene regulation in this system were demonstrated \cite{a26}. Also an increase level of noise allows switching more rapidly, but increases the rate of spontaneous switching and thereby decreases the precision of gene expression boundary \cite{a27}. Due to this trade-off, a proper level of intrinsic noise with respect to the extrinsic noise level is essential for robust and precise patterns \cite{a17}. This noise induce-switching is also found in mammalian embryo for maintaining cellular plasticity and organizing the blastocyst \cite{a28}. Intuitively, the underlying mechanism of noise-induced switching can be regarded as a battle between the extrinsic noise and the intrinsic noise. Such noise battle has also been reported in the stratified epithelial tissue maintenance: a balanced level of different noises is essential to homeostasis \cite{a29}. Cells may tune both intrinsic and extrinsic noise within certain ranges, then an optimal levels of these two noise can be achieved to get robust and precise spatial patterns.

\subsection{Cell-cell interactions to reduce variability in boundary formation}

The morphogen-mediated patterning relies on a long-range spatial signal, however, spatial patterns are fully sharpened within a short-range. It turns out gene expressions are not enough to fully explain the stratification and boundary formations. In the cellular level, cells movement driven by intercellular mechanical interactions is able to further refine the pattern.

To model the dynamics of multicellular systems including the mechanical intercellular interactions, the sub-cellular method is one cell-based modeling approach \cite{a31}. Consider a system with a constant number $N_{\text{cell}}$ of cells, each cell is composed of $M_{\text{node}}$ elements (nodes). For the $m$-th element in the $n$-th cell, the equation of motion for its position vector $\mathbf{x}_{n,m}$ takes form:
\begin{equation}
\centering
\frac{d}{dt}\mathbf{x}_{n,m}=\eta_{n,m}-\nabla_{\mathbf{x}_{n,m}}\sum_{j\neq m}^{M_{\text{node}}}\Phi_{\text{intra}}\left(\mid\mathbf{x}_{n,m}-\mathbf{x}_{n,j}\mid\right)-\nabla_{\mathbf{x}_{n,m}}\sum_{i\neq n}^{N_{\text{cell}}}\sum_{j}^{M_{\text{node}}}\Phi_{\text{inter}}\left(\mid\mathbf{x}_{n,m}-\mathbf{x}_{i,j}\mid\right).
\end{equation}
On the right hand side of the equation, the first term, $\eta_{n,m}$, is a Gaussian-distributed noise. The second and third terms represent intracelullar and intercellular interactions between the elements, respectively. These interactions are characterized by potentials $\Phi_{\text{intra}}$ and $\Phi_{\text{inter}}$. For example, the Morse potential allows repulsion and adhesion between two elements depending on their distance $r$:
\begin{equation}
\centering
\Phi\left(r\right)=U_0\text{exp}\left(-\frac{r}{\zeta_1}\right)-V_0\text{exp}\left(-\frac{r}{\zeta_2}\right).
\end{equation}
where $\left(U_0,V_0,\zeta_1,\zeta_2\right)$ are constants for each pair of elements and their values depend on the property (e.g. cell identity) of those cells they belong to. 

The selective intercellular adhesion, where cells with same/different identities have attraction/repulsion to each other, plays a role in the pattern formation (Figure \ref{fig_morphogen}D). This mechanism is regulated by effectors, such as cadherins or Eph receptors and ephrins, whose expression is related to cell identity \cite{a30}. Coupling cell-based models with gene expression models, the multi-scale hybrid models show that selective intercellular adhesions are crucial to sharpening boundaries in the segmental zebrafish hindbrain pattern formation \cite{a9} and stratification in skin epidermis \cite{a32}. Moreover, the asymmetric cell division contributes to better stratified level in epidermis than the symmetric division \cite{a33}.

\section{Conclusion and Perspective}

A never-fading topic in biology is how biological functions can be executed accurately and robustly. Noise attenuation, a general function required for many living systems, has drawn considerable attention of researchers. In this review, we first summarized many common network motifs for controlling extrinsic noise and discussed corresponding mechanisms especially the important role of timescale. For basic noise attenuation motifs, long linear cascades can buffer extrinsic noise because of its long response time, but accumulate intrinsic noise as the length of cascade grows. Other two classes of basic noise attenuation motifs are negative feedback loops and incoherent feedforward loops, whose noise buffering capabilities benefit from the low steady-state sensitivity. In systems where high sensitivity or switch-like response is required, positive feedback loops are preferred to attenuate noise due to the long response time and large SAT. For adaptive systems, sequentially coupling the noise attenuation module and the adaptation module, together with matching modules' response time, can mediate the trade-off between noise buffering capability and system's transient sensitivity, and thus favours the simultaneous implementation of adaptation and noise attenuation. However, a general principle for noise control especially the network dependence in oscillatory systems is still lacking. The relation between noise attenuation capability and network configurations helps to identify the complex regulatory networks in nature and provides guidance for synthesizing networks with robust biological functions. This will also be helpful for specific biological and medical applications.

While cellular processes can filter the noise to transfer accurate information, noise is also be exploited to achieve desirable functions.  A well-studied example is noise utilization in spatial patterns formation, which is also reviewed. To form a precise and robust readout from the noise morphogen gradient, noise attenuation is necessarily needed. Self-enhanced degradation contributes to robust morphogen gradient formation while the transcriptional network with an intermediate state benefits the in morphogen readout. Besides, a surprising result is that gene expression noise can be utilized to battle noise in morphogen, which leads to the sharpening of gene expression boundaries. Moreover, cell-cell interaction is shown to reduce variability in boundary formation.

Nevertheless, there are several open questions in the topic of noise control and utilization. First, while principles for buffering extrinsic noise have been well largely explored, how to control intrinsic noise by modulating network configurations remains unknown. Second, could we use machine learning to develop more efficient method to search for the principles for desirable functions, especially those integrating competing functions with the inherent trade-off? Third, can we apply the learned principles from regulatory networks to the spatial systems? Fourth, how cell-cell communication affects the cell decision in space? The investigations of these questions will be of great use.


\Acknowledgements{This work was supported by National Natural Science Foundation of China (11861130351, 11622102). QN is partly supported by a NSF grant DMS1763272, and The Simons Foundation grant (594598,QN).}



\bibliographystyle{unsrt}

\bibliography{ref}

\end{document}